\documentclass[conference]{IEEEtran}
\ifCLASSINFOpdf
\else
\fi
\usepackage{amsmath}
\usepackage{amssymb}
\usepackage{graphicx}
\usepackage{subfig}
\usepackage{algorithmic}
\usepackage{cite}

\newtheorem{proposition}{Proposition}
\newtheorem{theorem}{Theorem}
\newtheorem{definition}{Definition}

\begin{document}
%
\title{Cooperative Spectrum Sharing Between D2D Users and Edge-Users: A Matching Theory Perspective}



%


\author{
\IEEEauthorblockN{Yiling Yuan\IEEEauthorrefmark{1}, Tao Yang\IEEEauthorrefmark{1}, Yuedong Xu\IEEEauthorrefmark{1} and Bo Hu\IEEEauthorrefmark{1}\IEEEauthorrefmark{2}}
\IEEEauthorblockA{
\IEEEauthorrefmark{1} Research Center of Smart Networks and Systems, School of Information Science and Engineering\\
\IEEEauthorrefmark{2}Key Laboratory of EMW Information (MoE)\\
Fudan University, Shanghai, China, 200433\\
Emails: \{yilingyuan13, taoyang, ydxu, bohu\}@fudan.edu.cn}}

\maketitle

\begin{abstract}
The device-to-device (D2D) communication theoretically provides both the cellular traffic offloading and convenient content delivery directly among proximity users. However, in practice, no matter in underlay or overlay mode, the employment of D2D may impair the performance of the cellular links. Therefore, it is important to design a spectrum sharing scheme, under which the performance of both links can be improved simultaneously. In this paper, we consider the cell-edge user (CEU) scenario, where both sides have the demand to improve the quality of experience or service. Therefore, CEUs and D2D users both have intentions to form pairs, namely, CEU-D2D pairs, to cooperate mutually. Different from the conventional equilibrium point evaluation, the stable matching between D2D users and CEUs are formulated under matching theory framework instead. For each CEU-D2D pair, a two-stage pricing-based Stackelberg game is modeled to describe the willingness to cooperate, where the win-win goal is reached finally.
\end{abstract}


%
\IEEEpeerreviewmaketitle

\section{Introduction}
Recently, the wireless networks witness a dramatically increasing demand of local area service. In this context, a promising technology called device-to-device (D2D) communication, which enables direct communication between two mobile users in proximity without through base station, has attracted attention in both industry and academic \cite{22803, 5350367}. The adoption of D2D communications brings many advantages \cite{6163598}: allowing high-rate, low-delay, low-power transmission, extending the cellular coverage, etc.
\par
One big challenge for implementing D2D communication is how to allocate spectrum resource for D2D communications efficiently. Due to the controllable interference in licensed spectrum, it has been proposed that both the D2D and cellular users share the same spectrum, namely, underlay D2D and overlay D2D mode \cite{6805125}. For the former, the D2D and cellular links use the same spectrum at the same time, which could increase the spectrum reuse factor if a well designed interference management is available. For the latter, the operator allocates dedicated cellular resource to D2D links, which will incur lower spectrum efficiency albeit less interference.
\par
However, no matter in underlay or overlay mode, most literature mainly focuses on improving the performance of D2D links while ensuring that the performance of cellular links will not be severely degraded. The utility of cellular link is rarely considered. Furthermore, because information is directly exchanged between D2D users bypassing base station (BS),  the operator can only charge the D2D users based on how much resource they use irrespective of the data flow through D2D link\cite{6231164}, which may lead to lower utility. Therefore, it is still important for the operator to design a spectrum sharing scheme to improve its utility, which can also incentive D2D devices owned by selfish users to participate at the same time. On the one hand, for D2D users who aim to improve the quality of experience, the unlicensed band is free but too crowded, while sharing licensed spectrum provides higher performance but relies upon an agreement with BS. On the other hand, in cellular network, the cell-edge users (CEUs) usually suffer from poor channel condition so that their performance requirements are often hard to meet. Therefore, if D2D users assist CEU transmissions in exchange for access to licensed spectrum, the win-win outcome is achieved and higher benefit is available for the operators.
\par
Cooperative relay technology \cite{1246003} is a promising technology in improving the spectrum efficiency of cellular networks. In such scheme, the source broadcasts the signal to the destination and mobile users nearby first, and then these users help relay the received signal to the destination. The well-known relaying schemes include Amplify-and-Forward (AF) and Decode-and-Forward (DF). Motivated by this, many cooperative D2D communication relaying schemes for cellular networks \cite{6831680,6952682,7143335}, where D2D users can serve as a relay for the cellular user to earn opportunity to access the licensed spectrum band, are proposed. However, these schemes are designed from the perspective of D2D links.
\par
In this paper, we investigate a cooperative spectrum sharing scheme between D2D users and CEUs. When CEUs suffer from poor performance, they can select D2D transmitter as relay to satisfy their requirements, and these D2D pairs are able to access licensed spectrum in return. Utilities of cellular links and D2D links are both considered. A joint optimization framework based on game theory is proposed to characterize this kind of cooperation. Stackelberg game is used to model the interaction between CEU and D2D pair. Furthermore, when there are several CEUs and D2D pairs, they all seek appropriate partners to maximize their utilities. Thus we model the pairing problem as a marriage market to find stable CEU-D2D pairs given the preferences of both sides. Analytic and numerical results show that the proposed scheme can improve the performance of CEU and D2D users can gain considerable throughput, which makes both sides have intentions to cooperate. Moreover, under mild conditions, CEU can push the utility of the paired D2D user to be zero because of its leading role in the cooperation.
\par
The rest of this paper is organized as follows. In Section II, the system model and problem formulation are established. In Section III, The joint optimization algorithm is presented. The Section IV gives the numerical results and performance analysis, and finally Section V concludes this paper.

\section{System Model and Problem Formulation}
\subsection{System Model}
We consider a single cell of cellular networks. There are $M$ cell-edge users that communicate in conventional way through base station. We assume these users suffer from poor channel conditions so that their date rate requirements in the uplink can't be satisfied. Only uplink scenario is considered in this paper because of the limited power budget of user equipments. Let $\mathcal{M}=\{1,2,\cdots,M\}$ define the set of these users. Besides, there are $N$ transmitter-receiver pairs already operating in D2D mode. The set of D2D pairs is denoted by $\mathcal{N}=\{1,2,\cdots,N\}$. There is no dedicated channel for D2D communication. Therefore, in order to transmit its own data, the D2D user relays the cellular data in the uplink and gets access to the channels occupied by CEUs alternatively.
\par
\begin{figure}[!t]
\centering
\includegraphics[width=2in]{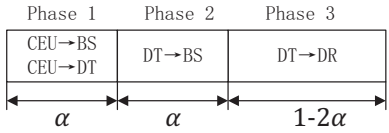}
\caption{Frame Structure}
\label{FrameStructure}
\end{figure}

We assume the time division multiple access (TDMA) technique is adopted. In case of cooperation, the normalized frame structure is shown in Figure.\ref{FrameStructure}. In order to reduce overhead, we assume only D2D transmitters are involved in relay procedure\footnotemark[1]. In the first phase, CEU broadcasts its data with power $P_{C}$ to D2D transmitter (DT) and BS. In the second phase, DT relays the received data with power $P_{D}$ to BS in decode-and-forward way. In the last phase, DT transmits its own data to D2D receiver (DR) with power $P_{D}$. The first phase and second phase both last $\alpha$ of the frame length, which constitute the relay transmission provided by D2D users for CEU. The third phase lasts $(1-2\alpha)$ of the frame length, which is allocated for D2D link. We refer to $\alpha$ as allocation coefficient.
\footnotetext[1]{The scheme proposed can also be applied to the case where D2D receivers are allowed to relay for CEU. In this case, for a D2D pair, the device bringing higher utility is selected to provide relay transmission for CEU, which, however, will introduce extra overhead.}
\par
The distanced based pathloss channel model considering multi-path fading\cite{6560489} is used in this paper. The notations for channel gains between different nodes are listed in table \ref{tab1}. We assume the channel gains are known to all nodes. Let $N_0$ denote the noise power.

\begin{table}[!t]
  \centering
  \caption{Notations for Channel Gains}
  \label{tab1}
  \begin{tabular}{|l|l|}
  \hline
  notation & physical meaning \\
  \hline
  $h_{ib}$ & channel gain between $i$th CEU and BS\\
  \hline
  $h_{ij}$ & channel gain between $i$th CEU and $j$th DT\\
  \hline
  $g_{jb}$ & channel gain between $j$th DT and BS \\
  \hline
  $g_j$ & channel gain between $j$th DT and $j$th DR\\
  \hline
  \end{tabular}
\end{table}

\subsection{Problem Formulation}
The achievable data rate of CEU $i$ in the direct link, denoted by $R^i$, is
\begin{equation}
\label{equ1}
R^i = \log_2(1+\frac{P_{C}h_{ib}}{N_0}),
\end{equation}
The outage refers to the case that $R^i < R_{th}$, where $R_{th}$ defines the data rate requirement.
\par
When cooperating with D2D user $j$, the data rate $R^i_{C}$ for $i$th CEU is limited by the minimum rate of the first two transmission phase, i.e.:
\begin{equation}
\label{equ2}
R^{ij}_{C}(\alpha_{ij}) = \alpha_{ij} \min \{ R_1^{ij},R_2^{ij}\},
\end{equation}
where
\begin{align*}
  R_1^{ij} & = \log_2(1 + \frac{P_{C}h_{ij}}{N_0}),\\
  R_2^{ij} & = \log_2(1 + \frac{P_{C}h_{ib}}{N_0} + \frac{P_{D}g_{jb}}{N_0}).\\
\end{align*}
For convenience, we define $r^{ij}_{C} \triangleq \min(R_1^{ij},R_2^{ij})$. If $R^{ij}_{C} \geq R_{th}$, the win-win situation will be achieved, which encourage CEU to cooperate with D2D pairs. At the same time, the achievable data rate of D2D link is
\begin{equation}
  \label{equ3}
  R^{ij}_{D}(\alpha_{ij}) = (1-2\alpha_{ij})\log_2(1 + \frac{P_{D}g_j}{N_0}) \triangleq (1-2\alpha_{ij})r^{ij}_{D},
\end{equation}
where $r^{ij}_{D}\triangleq\log_2(1 + {P_{D}g_j}/{N_0})$. However, if $R^{ij}_{C} < R_{th}$, the cooperation couldn't be reached and $R^{ij}_{D}=0$.
\par
Moreover, in order to avoid spectrum overuse, D2D pairs have to pay for using the licensed spectrum, referred to as spectrum leasing \cite{5457883}. Therefore, payoff function for D2D is defined as following:
\begin{equation}
\label{equ4}
\begin{split}
  U^{ij}_{D}(\alpha_{ij},c_{ij}) = & \beta_1 u_{D}(R^{ij}_{D}(\alpha_{ij})) - \beta_2P_{D}\alpha_{ij}- c_{ij}(1-2\alpha_{ij}).
\end{split}
\end{equation}
 In (\ref{equ4}), $u_{D}(\cdot)$ is the satisfaction of D2D pair $j$ with its data rate, $\beta_1$ is the equivalent revenue with respect to the satisfaction of D2D pair $j$ and $\beta_2$ is the cost per unit relay transmission energy. We define $u_{D}(\cdot)$ in the logarithmic form in this paper, namely $u_D(\cdot)=\ln(\cdot)$. Besides, $c_{ij}$ is the price coefficient. The payoff consists of three parts. The first term is the benefit from achievable transmission rate, the second term is the cost of energy for relay and the last term is the price charged for leasing spectrum.
 In addition, the payoff function of CEU can be defined as
 \begin{equation}
 \label{equ5}
   U^{ij}_{C}(\alpha_{ij},c_{ij}) = \beta_1 u_{C}(R_{C}^{ij}(\alpha_{ij})) + c_{ij}(1-2\alpha_{ij}),
 \end{equation}
 where $u_{C}(\cdot)$ is the satisfaction of CEU $i$ with its transmission rate. We define $u_{C}(\cdot)$ in the logarithmic form also. The first term is the equivalent revenue in terms of satisfaction and the second term is the payment from $j$th D2D pair. When $R^{ij}_{C} < R_{th}$, we define that $U^{ij}_{C} = U^{ij}_{D}=0$.
\par
Because all CEUs and D2D users are selfish, they have the incentive to maximize their performance. Therefore, it is natural to formulate the problem from a game theoretic perspective. In this paper, we focus on the following three problems.
\subsubsection{Pairing Problem}
Each CEU is selfish and can always make autonomous decision about its partner. Each D2D pair can also make its own decision to choose its partner. Therefore, it is important to decide a stable matching between CEUs and D2D pairs, in which each CEU and D2D pairs have no intention to deviate.
\subsubsection{Pricing Adjustment Problem for each CEU}
Each CEU can control the price charged to D2D pairs for access to its occupied channel.
\subsubsection{Spectrum Leasing Problem for each D2D pair}
Each D2D pair can decide the allocation coefficient $\alpha_{ij}$ to improve its performance further given the price imposed by CEU.
\par
\begin{figure}
  \centering
  \includegraphics[width=2.5in]{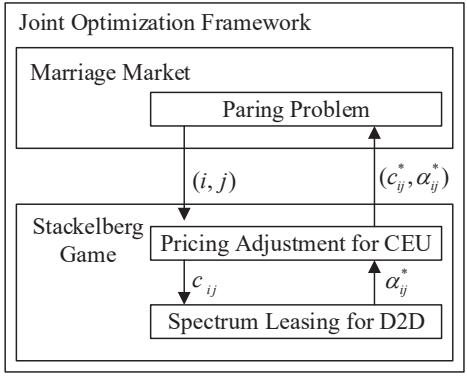}
  \caption{Joint optimization framework}
  \label{OptimizationFramework}
\end{figure}

Motivated by the hierarchical game model proposed in \cite{7244237}, we consider the joint hierarchical optimization framework addressing above three problem in this paper. More specifically, we model the pairing problem as marriage market problem. Then we establish a Stackelberg game, in which CEU is the leader to decide the price and D2D pair is the follower. The entire framework is depicted in Fig.\ref{OptimizationFramework}.

\section{Joint optimization Framework}
\subsection{Stackelberg Game for Pricing and Spectrum Leasing}
When CEU $i$ and D2D pair $j$ are paired, CEU $i$ can determine the price $c_{ij}$ charged to D2D pair $j$ in its occupied spectrum, and D2D pair can decide $\alpha_{ij}$ which would influence the utility of CEU. The interaction between CEU and D2D pair, that decides their actions to be taken in sequential manner, makes it natural to model the above pricing and spectrum leasing problem as a Stackelberg Game. The CEU is a leader and its action is to decide the price when D2D pair accesses its occupied spectrum. The D2D pair is a follower and it determines the $\alpha_{ij}$ according to the price imposed by CEU. We seek a Stackelberg equilibrium to the proposed problem using backward induction method.
\par
When given the price charged by $i$th CEU, the best strategy $\alpha^*_{ij}$ for $j$th D2D pair can be found as an optimization problem shown in (\ref{equ6}).
\begin{subequations}
\label{equ6}
  \begin{align}
      \max_{\alpha_{ij}}  & \ \ U^{ij}_{D}(\alpha_{ij},c_{ij}),\\
      \rm{s.t.}           & \ \ R^{ij}_{C} \geq R_{th},\\
                          & \ \ 0<\alpha_{ij}<0.5, \\
                          & \ \ R^{ij}_{C} = \alpha_{ij}r^{ij}_{C},
  \end{align}
\end{subequations}
The first constraint guarantees that CEU is willing to cooperate. It is easy to find out that if $r^{ij}_{C} \leq 2R_{th}$, the problem is infeasible.
\begin{proposition}
  Suppose $r^{ij}_{C} > 2R_{th}$, then the solution to problem (\ref{equ6}) is:
  \begin{equation}
  \label{equ7}
    \alpha^*_{ij}(c_{ij})= \begin{cases}
      \frac{R_{th}}{r^{ij}_{C}}, & c_{ij} \le \frac{\beta_2P_{D}}{2} + \frac{\beta_1 r^{ij}_{C}}{r^{ij}_{C}-2R_{th}}\\
      \frac{1}{2}-\frac{\beta_1}{2c_{ij}-\beta_2P_{D}}, & \text{otherwise}
    \end{cases}.
  \end{equation}
\end{proposition}
\begin{IEEEproof}
  Omitted for brevity.
\end{IEEEproof}
\par
As the leader of the game, CEU $i$ decides the price $c_{ij}$ to maximize its  utility with the knowledge of the strategy of $j$th D2D pair according to its decision. Therefore, the optimal price can be found as following:
\begin{subequations}
  \label{equ8}
  \begin{align}
      \max_{c_{ij}}  & \ \ U^{ij}_{C}\left(\alpha_{ij}^*(c_{ij}),c_{ij}\right)\\
      \rm{s.t.}           & \ \ U^{ij}_{D}\left(\alpha_{ij}^*(c_{ij}),c_{ij}\right) \geq 0,\\
                          & \ \ c_{ij} \geq 0,
  \end{align}
\end{subequations}
The first constraint is used to guarantee the D2D pair has the intention to cooperate. It is easy to verify that if $\beta_1\ln((1-2\frac{R_{th}}{r^{ij}_{C}})r^{ij}_{D})-\beta_2P_{D}\frac{R_{th}}{r^{ij}_{C}} < 0$, the problem above is infeasible.
\begin{theorem}
  Suppose the pair is formed by $i$th CEU and $j$th D2D pair, and below conditions are satisfied:
  \begin{align}
  \label{equ9}
    & r^{ij}_{C} > 2R_{th}, \\
    & \beta_1\ln((1-2\frac{R_{th}}{r^{ij}_{C}})r^{ij}_{D})-\beta_2P_{D}\frac{R_{th}}{r^{ij}_{C}} \geq 0.
  \end{align}
Then, $(c^*_{ij},\alpha^*_{ij})$ is the Stackelberg equilibrium of the game, where $\alpha^*_{ij}$ is given in (\ref{equ7}) and $c^*_{ij}$ is expressed as following:
\begin{equation}
  \label{equ11}
  c^*_{ij} = arg\max_{c\in\mathcal{C}} \ U^{ij}_{C}(c,\alpha^*_{ij}(c)).
\end{equation}
The set $\mathcal{C}$ is defined as following:
\begin{equation*}
  \mathcal{C} = \begin{cases}
     \{c_1,c_2\}                         &\underline{c}<c_2<\overline{c}\\
     \{c_1\}                             &\overline{c}<\underline{c}\\
     \{c_1,\overline{c},\underline{c}\}   &\text{otherwise}
  \end{cases},
\end{equation*}
where
\begin{small}
\begin{align*}
c_1 = & \min \left\{ \underline{c},{{{\beta_1}\ln ((1 - 2\frac{R_{th}}{r^{ij}_{C}})r^{ij}_{D}) - {\beta_2}\frac{R_{th}}{r^{ij}_{C}}{P_{D}}} \over {1 - 2\frac{R_{th}}{r^{ij}_{C}}}} \right\},\\
c_2 = & {{{\beta_2}{P_{D}}} \over 2} + {{{\beta_1}{\beta_2}{P_{D}}} \over {{\beta_2}{P_{D}} - 2{\beta_1}}}, \\
\overline{c} = &{{{\beta_2}{P_{D}}} \over 2} + {\beta_1}r^{ij}_{D}/{e^{1 + {{{\beta_2}{P_{D}}} \over {2{\beta_1}}}}},\\
\underline{c} = &{{{\beta_2}{P_{D}}} \over 2} + {{{\beta_1}r^{ij}_{C}} \over {r^{ij}_{C} - 2{R_{th}}}}.
\end{align*}
\end{small}
\end{theorem}
\begin{IEEEproof}
  Omitted for brevity.
\end{IEEEproof}

\par
If cooperation can't be reached, which means that problem (\ref{equ6}) or problem (\ref{equ8}) is infeasible, we define that $U^{ij}_{C}=U^{ij}_{D}=0$. Furthermore, under a mild condition, we find that $U^{ij}_{D}(c_{ij},\alpha_{ij})=0$.
\begin{proposition}
  If $\beta_2P_{D}<2\beta_1$, then $U^{ij}_{D}(c_{ij}^*,\alpha^*_{ij})=0$.
\end{proposition}

\begin{IEEEproof}
If $c_1 = \underline{c}$, then it means that the following inequality holds:
\begin{equation}
 \label{equ12}
  {{{\beta_2}{P_{D}}} \over 2} \leq {{{\beta_1}\log ((1 - 2\frac{R_{th}}{r^{ij}_{C}})r^{ij}_{D}) - {\beta_2}\frac{R_{th}}{r^{ij}_{C}}{P_{D}}} \over {1 - 2\frac{R_{th}}{r^{ij}_{C}}}}.
\end{equation}
After some algerbraic manipulations, we can get an inequality as following:
\begin{equation}
\label{equ13}
  {\beta_1}r^{ij}_{D}/{e^{1 + {{{\beta_2}{P_{D}}} \over {2{\beta_1}}}}} \geq {{{\beta_1}r^{ij}_{C}} \over {r^{ij}_{C} - 2{R_{th}}}}.
\end{equation}
Therefore, $\overline{c} \geq \underline{c}$. Besides, because $\beta_2P_{D}<2\beta_1$, we can find that $\frac{\text{d}U_{C}^{ij}(c,\alpha^*_{ij}(c))}{\text{d}c}>0$ over $[\underline{c},\overline{c}]$. Consequently, $U^{ij}_{D}(c_{ij}^*,\alpha_{ij}^*)=U^{ij}_{D}(\overline{c},\alpha_{ij}^*(\overline{c}))=0$.
\par
On the other hand, if $c_1 < \underline{c}$ ,it is easy to verify that $U^{ij}_{D}(c_1,\alpha_{ij}^*({c_1}))=0$ now. Using the similar idea, we can show that $\overline{c} < \underline{c}$. Therefore, we can also conclude that $U^{ij}_{D}(c_{ij}^*,\alpha_{ij}^*)=U^{ij}_{D}(c_1,\alpha_{ij}^*({c_1}))=0$.
\end{IEEEproof}
\par
Practically, $P_{D}$ is usually small, such as 0.1W. Moreover, the transmission power is not the major part of the power consumption for UEs, which means that $\beta_2$ is unlikely to be much larger than $\beta_1$. Therefore, the assumption in Proposition 2 is reasonable in most scenarios. So we assume the condition is met in our simulations. Intuitionally, Proposition 2 is resulted from the leading role of CEU in the cooperation. Although the utility is zero, D2D pair still has an intention to participate in the cooperation due to positive throughput.

\subsection{Matching Game for Pairing}
In this section, we study the pairing problem when there are several CEUs and D2D pairs. We will model pairing problem as a marriage market, also known as two-sided one-to-one matching market. Originally stemmed from economics\cite{Roth1992}, matching theory provides a mathematically tractable solution to handle with the problem of matching players in two distinct sets, according to each player's individual preference and information. Matching theory has become a promising framework for resource allocation in wireless communication.
This framework has many advantages\cite{7105641}: (1) It has efficient distributed implementations without centralized control; (2) Unlike most game-theoretic solutions, .e.g. Nash equilibrium, it has more suitable solution when applied to pairing problem; (3) It defines general preferences that can handle complex considerations.
\par
In our model, CEU and D2D pair can only be paired when they agree to cooperate with each other. Therefore, it is natural to model the interaction between the set of CEUs and the set of D2D pairs as an one-to-one matching game for solving the pairing problem. The CEU has a preference over all the D2D pairs. We use $\succ_i$ to denote the ordinal relationship of $i$th CEU. For instance, $j\succ_i j'$ means that $i$th CEU prefers $j$th D2D pair to $j'$th D2D pair. If $U^{ij}_{C}(c^*_{ij},\alpha_{ij}^*)>U^{ij'}_{C}(c^*_{ij},\alpha_{ij}^*)$, we have $j\succ_i j'$. Besides, if $U^{ij}_{C}(c^*_{ij},\alpha_{ij}^*) = U^{ij'}_{C}(c^*_{ij},\alpha_{ij}^*)$ and $R^{ij}_{C}\geq R^{ij'}_{C}$, we will have $j\succeq_i j'$. Similarly, we can define the preferences of D2D pairs. Let $\succ_j$ denote the ordinal relationship of $j$th D2D pair. We use the relationship $i\succ_i j$ to mean that agent $j$ is unacceptable to $i$, which is equivalent to the fact that agent $i$ and agent $j$ will not cooperate mutually in proposed scenario. Note that $i$th CEU is unacceptable to $j$th D2D pair if and only if $j$th D2D pair  is unacceptable to $i$th CEU.
\par
The major solution concept in matching game is \textit{matching} which is defined as follows.
\begin{definition}
  A \textbf{matching} is a function $\mu$ from $\mathcal{M}\times\mathcal{N}$ to $\mathcal{M}\times\mathcal{N}$ such that $\mu(m)=n$ if and only if $\mu(n)=m$, and $\mu(m)\in \mathcal{N}\cup \emptyset$, $\mu(n)\in \mathcal{M}\cup \emptyset$, for $\forall{n}\in\mathcal{N}$ $\forall{m}\in\mathcal{M}$.
\end{definition}
\par
The definition implies that the outcome matches the agent on one side to the one on the other side, or to the empty set. The agents' preferences over the matchings are coincident with their preferences over the matched partner in outcomes. In this paper, we seek a particular matching structure, which is defined as follows.
\begin{definition}
A matching $\mu$ is blocked by the CEU-D2D pair $(i,j)$, if $\mu(i)\neq j$ and $i\succ_j \mu(j)$, $j\succ_i \mu(i)$. A matching $\mu$ is individual rational if $\mu(i)\succeq_i i$ for $\forall i \in \mathcal{M} \cup \mathcal{N}$. A matching is \textbf{stable} if it is individual rational and not blocked by any CEU-D2D pair.
\end{definition}
\par
Deferred-Acceptance algorithm\cite{Gale1962} can be used to find a stable matching outcome. We propose an algorithm to solve the joint optimization problem, which is depicted in Algorithm 1. The algorithm consists of two stages. At the first stage, D2D pairs send their profile including channel state information(CSI) to the available CEUs. After receiving the information from D2D pairs, CEUs can calculate the price $c_{ij}$ according to Theorem 1 and  establish their preferences over D2D pairs respectively. Then CEUs send the information containing CSI and price $c_{ij}$ to D2D pairs. Each D2D pair can choose its best strategy $\alpha_{ij}$ according to the received information, and rank the CEUs depending on the achievable utility and data rate. At the second stage, each CEU proposes to its most preferred D2D pairs. Then each D2D pair will accept the most preferred one among the proposed CEUs and reject the rest. After that, the rejected CEUs propose to the next favourite D2D pairs and each D2D pair compares the new proposers and its temporary partner then selects the favourite one. The procedure will continue until no CEU is rejected. Any tie is broken in arbitrary way.

\begin{table}[t]
\label{alg1}
\begin{tabular}{p{0.9\linewidth}}
\hline
\textbf{Algorithm 1: A Joint Optimization Algorithm} \\
\hline
\textbf{Initialization}: Let $\mathcal{P}_i^{C}$ and $\mathcal{P}_j^{D}$ be the preference list of CEU $i$ and D2D pair $j$ respectively.\\
\textbf{Stage 1: Price Adjustment and Spectrum Leasing}
\begin{itemize}
\item[1]  CEUs and D2D pairs exchange their profile information:
    \begin{itemize}
    \item[i]   Each CEU $i$ computes the price $c_{ij}^*$ for every D2D pair $j$.
    \item[ii]  Each D2D pair $j$ chooses the best strategy $\alpha_{ij}^*$ according to $c_{ij}^*$.
    \end{itemize}
\item[2] Each CEU $i$ establishes its preference list $\mathcal{P}_i^{C}$ which only contains acceptable D2D pairs. And each D2D pair $j$ establishes its preference list $\mathcal{P}_i^{D}$, similarly.
\end{itemize}
\textbf{Stage 2: CEU-D2D Pairing}\\
WHILE $\exists m\in \mathcal{M}$ who was rejected
\begin{itemize}
\item[3]  Each CEU $j\in \mathcal{M}$ applies to its favourite D2D pair according to its preference.
\item[4]  Each D2D pair chooses the most preferred one considering the previous partner (if any) and the new applicants, and rejects the rest.
\item[5]  If CEU $j\in \mathcal{M}$ is rejected, it removes the D2D pair which it applies to at current round from its preference list $\mathcal{P}_j^{C}$.
\end{itemize}
\\\hline
\end{tabular}
\end{table}

\par
The complexity of stage 2 is $O(MN)$. The stable matching for marriage market always exists \cite{Roth1992,Gale1962}. Besides, we note that the outcome of Deferred-Acceptance algorithm is optimal for the set of players who make the proposals\cite{Roth1992}. Therefore, the proposed algorithm will lead to a stable matching between CEUs and D2D pairs, which is optimal in term of CEUs.
\par
It can be found that the stable matching structure is closely related to the preferences of CEUs and D2D pairs, which also depend on the resulting utilities from different CEU-D2D pairs formed. Furthermore, it turns out that the resulting payoffs are directly determined by allocation coefficients and pricing coefficients. By using the optimal pricing coefficients and allocation coefficients $(c_{ij}^*,\alpha^*_{ij})$, CEU $i$ and D2D $j$ can have the highest payoff when they form a CEU-D2D pair. Therefore, the outcome is stable in the sense that each CEU cannot improve its payoff further by unilaterally changing its price coefficient or paired partner and each D2D pair cannot improve its payoff, neither.

\section{Simulations}
The performance of our proposed joint optimization algorithm is investigated through simulation in this section. We use the pathloss based channel model considering multi-path fading\cite{6560489}. For example, the channel gain between CEU $i$ and BS can be expressed as:
\begin{equation}
\label{equ14}
  h_{ib} = K\gamma_{ib}L^{-\eta}_{ib}
\end{equation}
where $K$ is a constant determined by system parameter, $\gamma_{ib}$ is fast fading with exponential distribution, $\eta$ is the pathloss exponent and $L_{ib}$ is the distance between the BS and CEU $i$. The CEUs are distributed at the edge of the cell, and the D2D pairs are uniformly within the cell. We also assume the distance between the CEUs and DTs is less than 300 meters. When computing the utility function, we set $\beta_1=1$ and $\beta_2=10$. The detailed configuration parameters are depicted in Table.\ref{tab2}.
\begin{table}
\centering
\caption{Simulation Configure Parameters}
\label{tab2}
\begin{tabular}{|c|c|}
  \hline
  \textbf{Parameters}           &\textbf{Value}\\
  \hline
  Cell radius                   & 500\\
  Noise power($N_0$)            & -114dBm\\
  Pathloss constant($K$)        & $10^{-2}$\\
  Pathloss exponent($\eta$)     & 4\\
  CEU Tx power($P_{C}$)         & 100mW\\
  D2D Tx power($P_{D}$)        & 100mW\\
  Distance bewteen DT and DR    & 20m\\
  Numbers of CEUs ($N$)          & 20\\
  Required SNR for CEUs          & 5dB\\
  \hline
\end{tabular}
\end{table}

\par
\begin{figure}[!t]
\centering
\subfloat[Total utility of agents]{\includegraphics[width=1.6in]{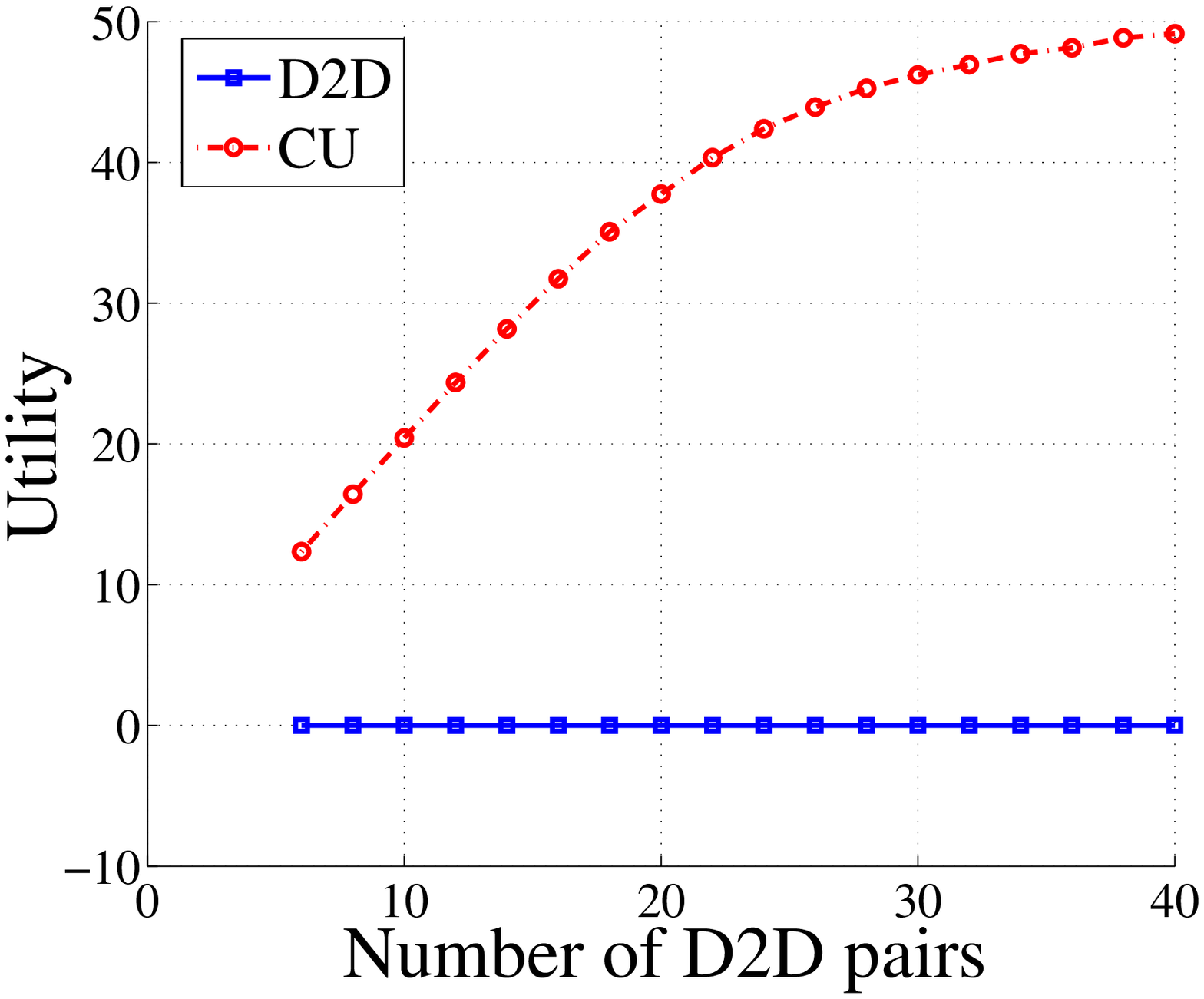}
\label{fig_first_case}}
\hfill
\subfloat[Sum-rate of agents]{\includegraphics[width=1.6in]{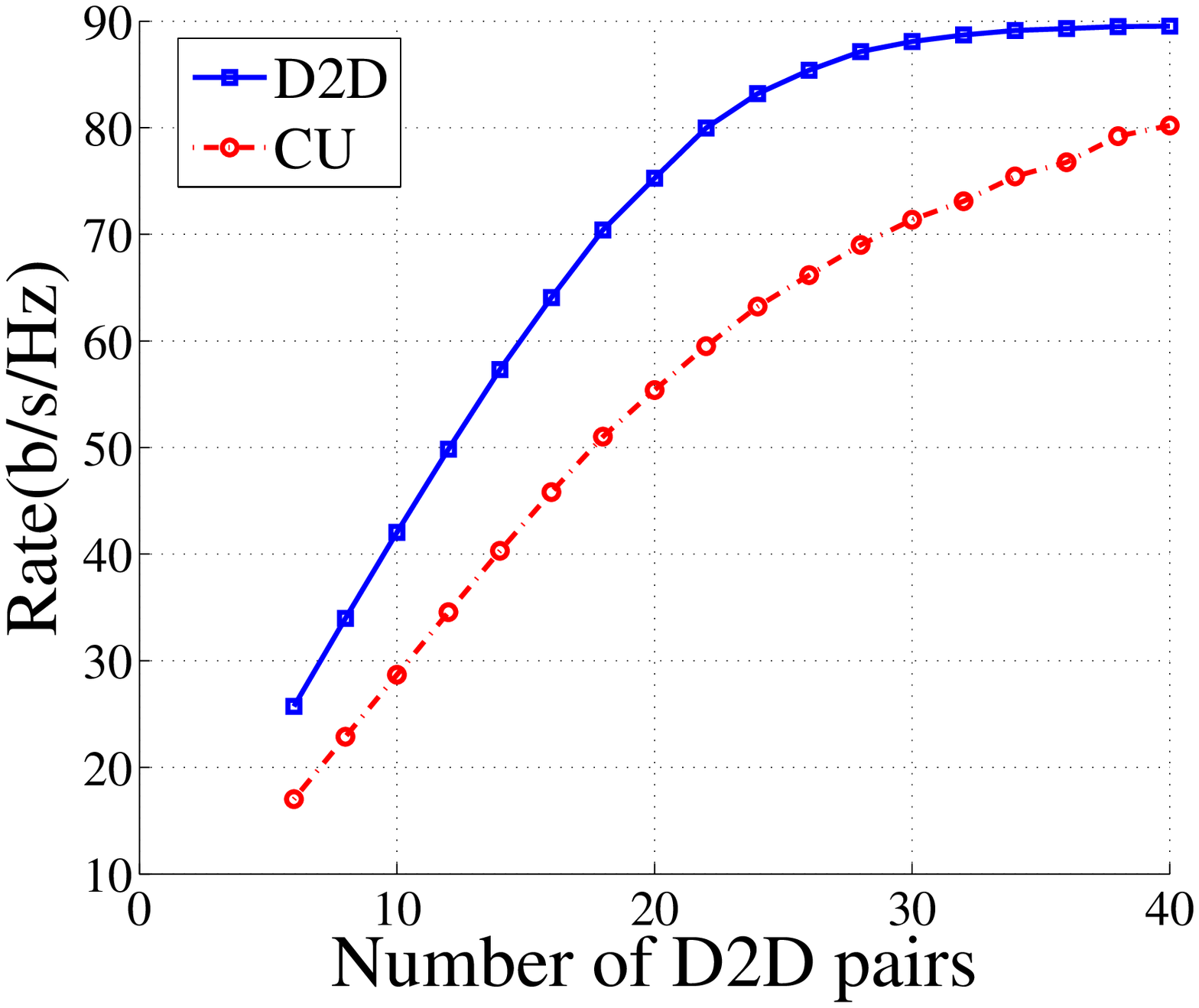}
\label{fig_second_case}}
\caption{Performance of proposed algorithm with different numbers of D2D pairs}
\label{fig_sim1}
\end{figure}

We first present the total utility and sum-rate of each agent with different numbers of D2D pairs using proposed algorithm in Fig.\ref{fig_sim1}. We can observe that the total utility of D2D pairs is always zero, which complies with Proposition 2. However, the zero-utility doesn't mean the zero sum-rate. The D2D pairs have high sum-rate although their total utility is zero. Consequently, D2D pairs are motivated to cooperate with CEUs. Besides, the total and sum-rate of CEUs are increasing with the increase of the number of D2D pairs. The major reason for performance improvement is that the more D2D pairs means the more opportunities to find better partners.

\par
We mainly compare the following two approaches for our scenario. The first one is \textit{random matching with Stackelberg game}. In this scheme, D2D pairs and CEUs are randomly matched. The optimal pricing coefficients and allocation coefficients are used if D2D pair and CEU are willing to cooperate mutually. The second one is \textit{stable random with fixed price}. In this scheme, D2D pair decides the best strategy given fixed price coefficient. Afterwards the second stage of Algorithm 1 is used to decide the matching between D2D pairs and CEUs. The proposed algorithm is referred to as \textit{stable random with Stackelberg game}. We will focus on the performance of CEUs, because of their leading roles in licensed spectrum.

\begin{figure}
  \centering
  \includegraphics[width=2.7in]{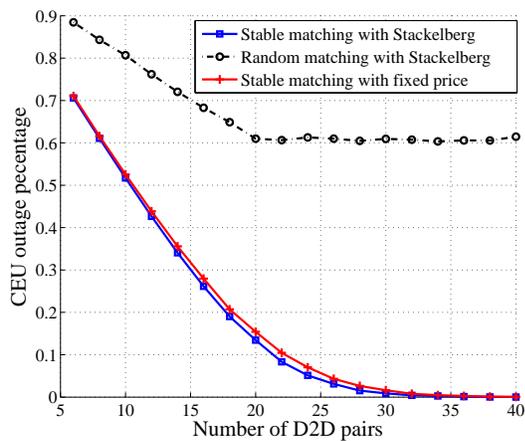}
  \caption{Outage percentage of CEU with different numbers of D2D pairs}
  \label{per5}
\end{figure}
\par
Fig.\ref{per5} compares the outage percentage of CEUs with different schemes for different number of D2D pairs. Our proposed algorithm has the least outage percentage. Moreover, we can observe that the performance achieved by stable matching has significant gain over that by random matching. That's because stable matching takes the preference of each agents into consideration while the random matching doesn't. Because outrage occurs only when matched partners are unacceptable to each other and thus price adjustment has a little effect on outrage percentage, our scheme performs a little better than the stable matching with fixed price. In addition, when the number of D2D pairs is more than 20, the outrage percentage achieved by random matching remains unchanged. That's because the number of CEUs is 20, and the available D2D pairs is not enough when the number of D2D pairs is less than 20. In this situation, more D2D pairs means that more CEUs can be matched which leads to the decrease of outrage percentage. However, when there are more than 20 D2D pairs, each CEU have a matched partner, and outrage percentage will stay almost unchanged because of random matching.

\begin{figure}[!t]
  \centering
  \includegraphics[width=2.7in]{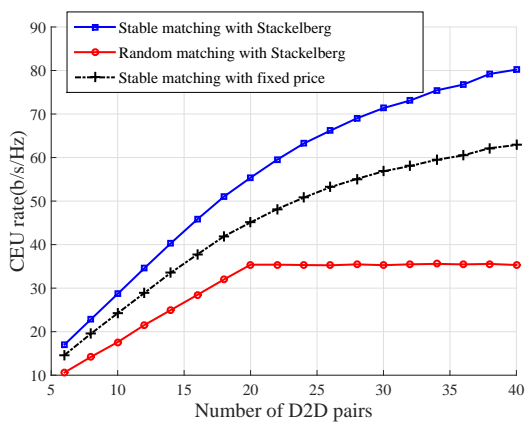}
  \caption{Sum-rate of CEU  with different numbers of D2D pairs}
  \label{per3}
\end{figure}
\par
Fig.\ref{per3} compares the sum-rate of CEUs achieved by three schemes with different number of D2D pairs. As the number of D2D pairs increases, CEUs get more chance to access the cooperating relay and sum-rate will be improved. Fig.\ref{per3} shows that the proposed joint optimization algorithm yields considerable gain over other schemes. Besides the benefit of stable matching, price adjustment can improve the performance of CEU further. Because of the same reason we have mentioned, the performance of random matching increases at first and remains unchanged when there are enough D2D pairs.

\section{Conclusion}
In this paper, we investigate a cooperative spectrum sharing scheme between D2D users and CEUs, where D2D user relays the cellular data in the uplink to get access to the licensed channel, so that both sides can improve the quality of service through cooperation. Thus, unlike underlay and overlay mode, a win-win situation is achieved. Given the selfishness of each sides, we use a Stackelberg game to describe the willingness to
cooperation. Further, we establish a stable marriage market to study the paring problem. We present numerical results to verify the efficiency of the proposed scheme.

\section*{Acknowledgment}
This work was supported by the National Science and Technology Major Project of China (Grant No. 2015ZX03001040-002) , the NSF of China (Grant No. 61501124, No. 61402114)




%


\bibliographystyle{IEEEtran}
\bibliography{IEEEabrv,reference}

\end{document}